\documentclass[conference]{IEEEtran}
\IEEEoverridecommandlockouts
\usepackage{cite}
\usepackage{amsmath,amssymb,amsfonts}
\usepackage{algorithmic}
\usepackage{graphicx}
\usepackage{textcomp}
\usepackage{xcolor}
\usepackage{subcaption}
\usepackage{booktabs} 
\usepackage{graphics}
\usepackage{listings}
\usepackage{color}
\usepackage{epstopdf}
\usepackage{epsfig}

\definecolor{dkgreen}{rgb}{0,0.6,0}
\definecolor{gray}{rgb}{0.5,0.5,0.5}
\definecolor{mauve}{rgb}{0.58,0,0.82}

\begin{document}
\title{Ownership preserving AI Market Places using Blockchain}
\author{\IEEEauthorblockN{Nishant Baranwal Somy} 
\IEEEauthorblockA{\textit{IIT Kharagpur, India}\\
  somy1997@gmail.com  
}
\and
\IEEEauthorblockN{Kalapriya Kannan}
\IEEEauthorblockA{\textit{ IBM Research, India  } \\
\textit{kalapriya.kannan@in.ibm.com }
}
\and
\IEEEauthorblockN{Vijay Arya}
\IEEEauthorblockA{\textit{ IBM Research, India} \\
\textit{ vijay.arya@in.ibm.com}}
\and
\IEEEauthorblockN{Sandeep Hans}
\IEEEauthorblockA{\textit{IBM Research, India} \\
\textit{shans001@in.ibm.com }}
\and
\IEEEauthorblockN{Abhishek Singh}
\IEEEauthorblockA{\textit{IBM Research, India} \\
\textit{ abhishek.s@in.ibm.com}}
\and
\IEEEauthorblockN{Pranay Lohia}
\IEEEauthorblockA{\textit{IBM Research, India} \\
\textit{plohia07@in.ibm.com}
}
\and
\IEEEauthorblockN{Sameep Mehta}
\IEEEauthorblockA{\textit{IBM Research, India} \\
\textit{sameepmehta@in.ibm.com}
}
}

\maketitle

\begin{abstract}
We present a blockchain based system that allows data owners, cloud vendors, and AI developers to collaboratively train machine learning models in a trustless AI marketplace. Data is a highly valued digital asset and central to deriving business insights. Our system enables data owners to retain ownership and privacy of their data, while still allowing AI developers to leverage the data for training. Similarly, AI developers can utilize compute resources from cloud vendors without loosing ownership or privacy of their trained models. Our system protocols are set up to incentivize all three entities - data owners, cloud vendors, and AI developers to truthfully record their actions on the distributed ledger, so that the blockchain system provides verifiable evidence of wrongdoing and dispute resolution. Our system is implemented on the Hyperledger Fabric and can provide a viable alternative to centralized AI systems that do not guarantee data or model privacy. We present experimental performance results that demonstrate the latency and throughput of its transactions under different network configurations where peers on the blockchain may be spread across different datacenters and geographies. Our results indicate that the proposed solution scales well to large number of data and model owners and can train up to 70 models per second on a 12-peer non optimized blockchain network and roughly 30 models per second in a 24 peer network. 
\end{abstract}

\section{Introduction}
 A number of AI Marketplaces \cite{singularitynet, atmatrix,burstiq,neureal,deepsee,fysical}
are being set up as 
collaboration hubs to enable different stakeholders
 in the AI value chain to connect, develop, and monetize AI
  assets i.e. data and models, in a secure manner. Additionally, 
  these marketplaces aim to accelerate innovation, promote responsible 
  use of AI, and fair distribution of value generated by 
  the development and use of AI assets. For example, consider
   a group of $k$ hospitals each of whom have a certain amount of 
   patient healthcare data, but lack expertise to jointly build AI models.
    At the same time, AI developers in academia and industry generally 
    lack access to patient healthcare data. In this setting, an AI marketplace
     can enable collaboration between the hospitals (i.e. data owners)
      and AI developers to securely build models to assess patient health. 
      Moreover, cloud vendors can contribute \textsc{gpu} compute resources 
      to train these models and any model building blocks can be reused by other
       developers. Finally, these models can be discovered and consumed by 
       businesses including clinics, insurance, and pharmacy companies.

A critical factor impeding the success of both centralized and decentralized AI marketplaces is that
 they do not guarantee data and model privacy~\cite{algorithmia, ocean, kaggle}.
 As a consequence, both data and model owners can easily lose ownership of their assets and are unable to derive value from them in a sustainable manner. 
 Moreover, large scale sharing of data or models may not be feasible due to regulatory 
 constraints \cite{gdrp} and also because owners might lose competitive intellectual property and economic advantage. 
 Additionally, centralized AI marketplaces are dependent on a trusted central 
 entity to maintain a verifiable audit trail of data sharing and training, 
 which has the potential to create digital monopolies, increase costs, 
 and is open to malpractice.

%



This work presents the design and architecture of a blockchain based solution that preserves the privacy and ownership of AI assets in a decentralized AI marketplace that 
 has no trusted central entity. Our system considers three classes
  of market participants: data owners (DO), model developers or owners (MO) 
  and cloud owners (CO) and allows them to collaboratively 
  train AI models on available datasets using federated learning~\cite{45648}. 
  In our system,  data privacy is ensured by splitting each
   dataset across multiple COs so that no single entity on the 
   blockchain has access to the entire dataset. 
   Each CO that holds a data subset then participates in multiple rounds of training using 
   federated learning in order to build the model.   
   Model privacy 
   is guaranteed by training models that are encrypted using fully
    homomorphic encryption, so that model predictions are unusable without 
    the decryption key~\cite{8260844}. Our system has been implemented using 
    the open source Hyperledger Fabric \cite{hyperledger} wherein
     all stakeholders interact with the system through chaincode
      functions (equivalently smart contracts on Ethereum network). 
      We present the design of these functions, which incentivize truthful 
      recording of all transactions on the
       blockchain including splitting and distribution
        of datasets and the scheduling and execution of
         multiple rounds of training across COs. This ensures that the system
          provides verifiable evidence of expected behavior or wrongdoing and 
          dispute resolution, thus building trust with all stakeholders. For
           instance, the system allows an MO to easily verify that 
           the data as proposed by the DO is indeed the
            data used to train her model. Similarly, an MO can verify 
            that each CO participating in federated learning has indeed
             submitted a unique intermediate model based on a round 
             of training on its data, as opposed to submitting a copy of a 
             trained model from another CO.

We have deployed our solution across a blockchain network composed of multiple
 organizations spread across three different locations
  with each organization contributing up to $24$ peers. We 
execute transactions related 
to collaborative training of AI models and present experimental performance
 results that demonstrate the latency and throughput of these transactions under different
  network configurations. Our experiments show that our system scales 
  well and can support training of up to $70$ models/second 
  with sub second latencies observed for recording information in the blockchain. 

The rest of the paper is organized as follows. Section~\ref{sec:systemoverview} presents the design and architecture of our system including details of its chaincode functions.
 Section~\ref{sec:experiments} presents experimental evaluation results, followed by sections~\ref{sec:relatedwork} and ~\ref{sec:conclusions}, which present related work and conclusions respectively. 

\section {System Overview }
\label{sec:systemoverview}


Privacy preserving environments for data sharing and model training when data, models, and compute resources are offered in a trustless setting, require both
efficient protocols and platforms capable of truthfully recording and verifying the sequence of actions by different stakeholders.

\begin{figure}
  \centering
  \includegraphics[scale=0.3]{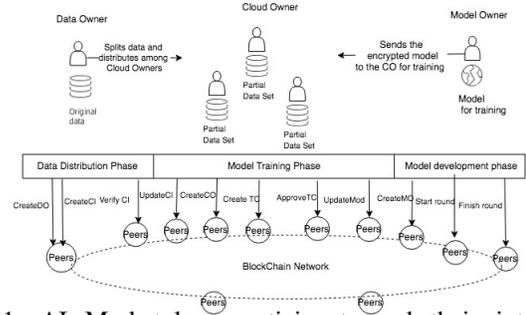}
  \vspace{-3mm}
  \caption{\label{figure:blockchain-interactions}AI Marketplace participants
   and their interaction with Blockchain via different chaincode functions 
   (i.e. smart contracts). We have implemented 15 chaincode functions to interact and store
   information in the blockchain.  }
\vspace{-5mm}
\end{figure}

\noindent{\bf Stakeholders. } Our system models three marketplace stakeholders: \emph{data owners} $(DO)$, \emph{cloud owners} $(CO)$, and \emph{model owners} $(MO)$. The $DO$s own large proprietary datasets (e.g. healthcare, self-driving cars, compliance data, etc.), which are especially valuable to train accurate AI models. They wish to monetize this data and sell it for AI training in a safe, secure, and transparent manner multiple times without loosing its ownership. By being part of a marketplace, $DO$s can increase the outreach and monetary gains that they can derive from their data, which may otherwise be utilized minimally or lie unused in their datalakes. 
The $CO$s are cloud service providers, who wish to sell storage and GPU compute resources needed to train AI models. By being part of a marketplace, the COs can increase their customer base and offer subscription-based storage and compute services at competitive prices. 
The $MO$s are enterprise or freelance AI developers who have the skills and experience to develop sophisticated AI models, but lack the data needed to train the models. The MOs eventually wish to monetize their trained models and therefore do not want to loose ownership of their models in the process of training them. By being part of a marketplace, MOs can obtain access to diverse datasets that meet their training requirements, GPU compute resources at competitive prices from COs, and ultimately lower the overall cost of training AI models. 
Figure \ref{figure:blockchain-interactions} shows the different marketplace stakeholders, each of whom is motivated to participate in the marketplace based on their personal economic gains.

 
\noindent{\bf System Operation. } There are two distinct phases of system operation for training AI models so that both the DOs and MOs retain privacy of their assets, i.e, Data distribution phase and Collaborative model training phase.  

\vspace{0.5mm}\noindent\emph{Data distribution phase. }The protocol begins with $DO$s expressing their intent to share their datasets in the marketplace for AI training. However, any exposure of the whole dataset can lead to data leaks to other parties in the system. In order to preserve the privacy of their datasets, the $DO$s can potentially share encrypted datasets, in which case AI training would need to be performed on encrypted data. However when MOs wish to monetize trained models, model inferencing would again require data to be fed in encrypted form. This implies that MOs may need to depend on DOs for encryption of new input data for inference or possibly decrypting the prediction response. Our system uses an alternative novel approach wherein the dataset is split into multiple subsets and stored across COs so that no single CO has full access to the dataset and each CO holds a small fraction of the overall data. The $DO$ enters into an offline contract with a set of $CO$s and securely transfers the data subsets off-chain to the individual $CO$'s. The COs record the receipt of data subsets on blockchain, which is verified and acknowledged by the DO. While recording information on blockchain, each entity uses pseudo identifiers for anonymization and the actual identities of entities is not revealed. Thus only the DO knows the identities of all COs that store its data and none of the COs know each other.

\vspace{0.5mm}\noindent\emph{Collaborative training phase. } Once a dataset is distributed to the $CO$s, it is available for purposes of AI training in the marketplace. We refer to these data subsets as privacy preserving data subsets (PPDS). An MO develops a model with the help of data samples that DOs expose and obtains permission to train the model on the entire dataset. Since the dataset is distributed across multiple COs, our system employs federated learning to train the AI models, wherein each CO contributes compute resources towards training. However, in federated learning, all COs obtain access to the final trained model, which implies that MOs would loose ownership of their models during the training phase. In order to avoid this, the MOs encrypt their models using homomorphic encryption and the training proceeds in rounds. During each round, the MO shares the running version of the encrypted model with all COs. Each CO trains the encrypted model using its own data subset and returns back the trained model to the MO. The MO aggregates the individually trained models are shares the updated version with the COs for a next round of training, eventually obtaining a final trained encrypted model. Training of homomorphically encrypted models is a well studied area \cite{homo1} \cite{homo2} \cite{homo3}. We do not delve into the details of the model encryption but choose an approach that seamlessly blends homomorphic encryption with federated learning. 

Additionally, since $MO$s have access to partially trained models at the end of each round, they can potentially learn characteristics of data stored by each CO. In order to avoid this, during each each round, a random set of COs holding a dataset are utilized for training. In this manner, while all data subsets are eventually utilized for training over multiple rounds, it becomes difficult for $MO$s to decipher the data characteristics of individual COs, thereby guaranteeing complete integrity of data ownership.

There are a number of challenges in realizing the above steps in a trusteless setting using blockchain. Specifically, 
\textit{how does one ensure that the market operation is transparent to all parties? 
How can one build a trusted platform for data sharing and collaborative training such that participants can record actions without exposing data and models? 
How can the protocol ensure that parties cannot collude with each other?
How can all parties record their actions such that the system automatically provides evidence of expected behavior or wrongdoing and dispute resolution?}

Our system leverages Blockchain to enable all stakeholders to truthfully
 record the sequence of events during data sharing and training. Blockchain has 
 the advantage of enabling trust between different non-trusted entities in
 a marketplace. For details of how blockchain works, one can refer 
 to \cite{hyperledger} that provides a comprehensive overview of
  an open source blockchain implementation, the Hyperledger Fabric. 
  To provide complete transparency and preserve the ownership 
  in a trusted manner our system uses blockchain for
   recording and validating all operations.  We assume that all 
   stakeholders subscribing to the protocol will either contribute nodes
    to the blockchain network to facilitate their requests and integrate
     the APIs to interact with the blockchain for all required events as 
     specified in the protocol. In this paper, we turn our attention to designing the system and protocol 
     that would enable all the
     stakeholders to participate in AI training without the fear of losing ownership.
      With our system, existing techniques
     for model encryptions and data splitting can be easily plugged in. Therefore, we delve less
     on individual mechanisms of optimized data splitting or model encryptions
     and focus completely on designing
     a system that would enable ownership preservation.

Existing works like \cite{algorithmia} \cite{SkyChain} uses the blockchain
 to both store data and train models.  However, our system decouples the 
 storage of data and models from the blockchain (ie., peers in the underlying 
 blockchain network do not store data or train models). This design brings in 
 twin benefits. Firstly, unlike \cite{algorithmia} the network nodes need not 
 perform redundant model training operations to provide proof of work or 
 achieve consensus. This eliminates the computation and storage overhead 
 from network nodes. Secondly, the communication with blockchain can be designed as shorter messages using blockchain for recording purposes only. 
 This naturally enables existing data and model platforms to plug-in and integrate into any blockchain network seamlessly. Figure~\ref{figure:blockchain-interactions} 
 shows the interactions of the participating parties with the blockchain network. The interactions are defined in the subsequent sections.

\subsection{The protocol and Interaction with Blockchain}
In this section we walkthrough the protocol and interaction with the blockchain. 
Figure \ref{figure:blockchain-interactions} 
illustrates the interaction of different stakeholders with the blockchain.  As mentioned earlier,
we assume that the different stakeholders will use our API's to truthfully record the
events in the blockchain. Our design ensures that blockchain performs a few critical 
validations and verification and the stakeholders will not be able to proceed without
these validations.  This makes blockchain relevant and bypassing the blockchain interactions would not be feasible.
\subsubsection{Blockchain members and assets}

\begin{figure}
  \centering
  \includegraphics[scale=0.25]{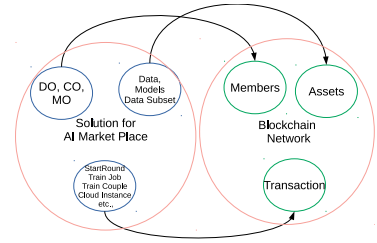}
  \caption{Mapping between solution and blockchain based entities \label{fig:mapping}}
  \vspace{-0.2in}
\end{figure}

As illustrated in figure~\ref{fig:mapping}, our solution for AI marketplace leverages capabilities of the underlying blockchain network to define members, assets, and transactions. 
Members are essentially participants or stakeholders in the AI marketplace which includes $CO$s, $MO$s, and $DO$s. 
Real world resources are modeled as assets in the blockchain, which in our case includes the Data, Data Subsets, and Models.  
Transactions (or chaincode functions) enable members to perform a set of predefined operations on the assets related to collaborative training. 
Member access to perform operations on assets is controlled by the access control list (\textsc{acl}). 




New members are created through the Create calls. 
A sample representation of $DO$ in the blockchain is illustrated in the listing \ref{lst:do} (members are treated as assets in the blockchain). 
For example, \emph{createDO} call returns a unique identifier and allows a $DO$ to register and join the marketplace. 
Similarly other create calls are used to create $MO$ and $CO$ entities and their identifiers are recorded in blockchain. 
\begin{center}

\begin{tiny}
  \noindent\begin{minipage}{.2\textwidth}
    \begin{lstlisting}[caption= $DO$ definition,frame=tlrb, label={lst:do} ]{CloudInstance metadata}
asset DO identified by id {
  o String id
  o String name optional
  o String organization optional
  o String howMany optional
}     
    \end{lstlisting}
  \end{minipage}
\end{tiny}
\end{center}


Our solution defines 10 different types of assets to store different entities and their states. 
We enumerate a few critical assets and describe in detail their purpose throughout the protocol.  
Assets are essentially represented as data structures within blockchain and used by $CO$s, $MO$s and $DO$s to record information during different steps of collaborative training. 
For instance, corresponding to each $CO$, there exists a an asset called \emph{Cloud Instance} (CI), which links the CO to the Data chunk it holds (including data subsets and their replicated counterparts). 
Three main components in the $CI$ are the subset ID, the $CO$ identifier, and a field that tracks the status of the Data Subset. 
It should be noted that the actual identifiers of $CO$s, $MO$s, and $DO$s (i.e. an ip address or a URL), is never actually stored on the blockchain.  Therefore, no parties can query the blockchain and obtain the identifier to establish a direct contact with $CO$.  The blockchain provides a pseudo-random identifier for every member, which is used to store and refer to the members for purposes of querying.

Another asset called \emph{Train Couple} ($TC$) is used to represent a model training instance, which associates a model with a data subset. A $TC$ has a $DSS$ and a model object 
as its member variables. The status field indicates whether the model is ready to be trained or has completed the training phase. The listings \ref{lst:ci} and  \ref{lst:tc} show the  $CI$ and $TC$ asset definitions respectively as represented in the blockchain.

\begin{tiny}
  \noindent\begin{minipage}{.2\textwidth}
  
    \begin{lstlisting}[caption= $CI$ metadata,frame=tlrb, label={lst:ci} ]{CloudInstance metadata}
      asset CloudInstance 
      identified by id {
        o String id
        --> CO co
        --> DSS dss optional
        o CIStatus status optional
        o String nonce optional
        o String hash optional
        o Integer rounds optional  
      }                
    \end{lstlisting}
    \end{minipage}\hfill
    \begin{minipage}{.2\textwidth}
    \begin{lstlisting}[caption=$TC$ metadata,frame=tlrb, label={lst:tc}]{TrainCouple metadata}
      asset TC identified by id {
        o String id
        --> DS ds
        --> Mod mod   
        o TCStatus status
        o Integer rem                       
        o Boolean paid
        o Integer round   
      }
    \end{lstlisting}
    \end{minipage} 
  \end{tiny}

Similar to assets, our solution includes about 37 different transactions for enabling collaborative training between different stakeholders in a trustless setting. About 15 of these are called directly by different members while the remaining ones are called from within other transactions. We enumerate a few critical transactions below. 
  
The \emph{StartRound} transaction (listing \ref{lst:sr}) is used by the $MO$s to begin the training process wherein blockchain chooses a random set of data subsets which are trained by each cloud owner independently for federated learning. 

\begin{tiny}
  \begin{lstlisting}[caption={Chaincode for StartRound Transaction}, label={lst:sr}]
    func (t *DataMarketChaincode) StartRound(stub shim.ChaincodeStubInterface, args []string) peer.Response {
    //Initiate a TC
	var tc TC
	tcid := args[0]
	tcBytes, err := stub.GetState(tcid)
	if tcBytes == nil {
		fmt.Printf("tc with id %s do not exists\n", tcid)
		return shim.Error("tc do not exists")
  }
  
  //Check if the TC is approved by the DO so that model training can start.
	if tc.Status != "APPROVED" {
		fmt.Printf("TC with id %s not yet approved\n", tcid)
		return shim.Error("tc not yet approved")
  }
  //Track the number of rounds in the Blockchain
  tc.Round += 1
//Get random list of DSS to train in each round and initialize the `rem' to the number. Each $CO$ updates rem by subtraction when its training is complete
	tc.Curdssidlist = nil
	var r int = (tc.Round+1) % 2
	var ds DS
	dsid := tc.Dsid
	dsBytes, err := stub.GetState(dsid)
	tc.Rem = 0
	for i := 0; i < len(ds.Dssidlist); i++ {
		var dss DSS
		dssBytes, err := stub.GetState(ds.Dssidlist[i])
		if i%2 == r {
			tc.Curdssidlist = append(tc.Curdssidlist, dss.Id)
			tc.Rem += len(dss.Ciidlist)
		}
	}
  fmt.Printf("Going to store tc\n")
  //Store the states in blockchain
	err = stub.PutState(tcid, newTcBytes)
	return shim.Success(newTcBytes)
}
    \end{lstlisting}
  \end{tiny}

For sake of brevity we skip other transactions but describe them in the context of protocol in the subsequent sessions.

\subsection{Data Distribution and Acknowledgement in Blockchain}

One of the first steps involved collaborative training is the splitting and replication of a dataset by the $DO$ among different $CO$s. In our solution, both these steps are performed intelligently to ensure that the ownership of the dataset is preserved and that malicious $CO$s do not falsely claim to have trained their models using their data subsets, without actually training them. 

\noindent\emph{Data Splitting}. As explained previously, the $DO$ makes an off-chain agreement with a certain number of $CO$s and sends the data subsets to them securely. In our solution, no information about the actual dataset itself is recorded on the blockchain to ensure its privacy. The $DO$ splits the dataset such that each $CO$ has a small subset of the entire dataset. Moreover, the splitting is done in a skewed fashion i.e. no subset has all the classes of the data or range of values. Also, no subset has majority of the data from a single class. The first condition ensures that no single $CO$  can derive meaningful information about the dataset, while the second condition ensures that using multiple queries or using multiple model iterations, an $MO$ cannot interpret the data characteristics. 

In terms of steps recorded on blockchain, the $DO$ creates a $CI$ instance by using the transaction `CreateCI' for each chunk of data that gets distributed. The initial status is set to `Free` and all the $CO$s are notified. The $CO$s actively wait for events on blockchain and upon receiving a notification, each $CO$ prepares itself to receive the data offchain from the $DO$. Thus each $CO$ knows the identity of the $DO$ and vice-versa, however none of the $CO$s know each other. 

A fraudulent $CO$ can claim to have received a different data subset than the one provided by the $DO$.  In order to avoid this scenario, our protocol requires that the $CO$ declare the hash of the data subset it receives on blockchain.  The $CO$ records the hash values of its data subset in the asset $CI$. It uses the transaction $JoinCI$ to update the hash value on blockchain. The $DO$ consults the blockchain and verifies the hash declared by the $CO$ against its own computation of the hash on the data subset that it distributed to the $CO$. If there
 is a mismatch, the $DO$ can tag the $CO$ as fraudulent. The $DO$ uses the transaction $VerifyCI$ to update the status in the $CI$ to `verified` if the $DO$ sees a matching hash, otherwise it remains set as non-verified. This step ensures that $CO$ receives no faulty data subset. Once a $DO$ verifies the hash of a $CO$'s data subset and marks the $CI$ status as verified, the data chunks become ready for training. 
 
\begin{figure}[h!]
  \centering
  \includegraphics[scale=0.3]{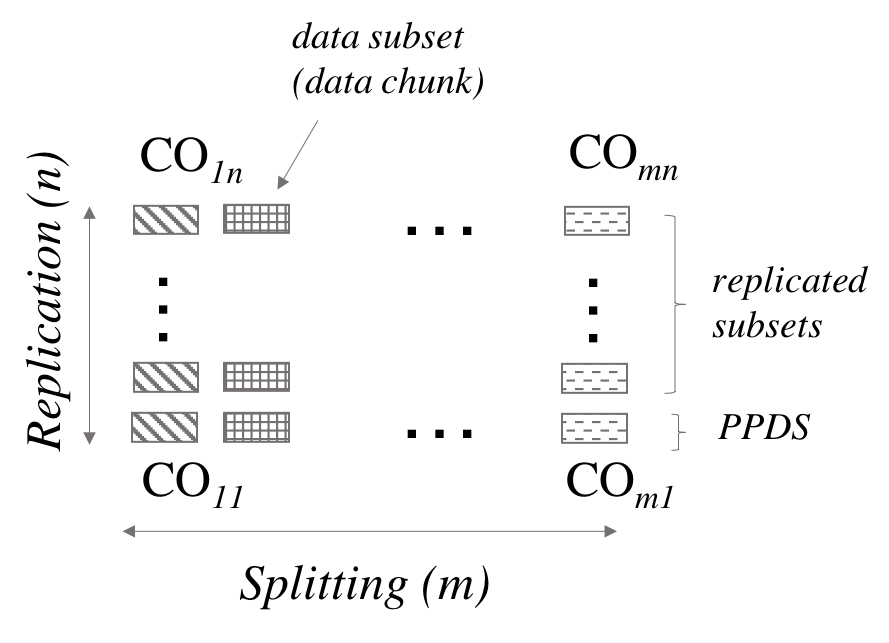}
  \vspace{-3mm}
  \caption{\label{figure:datadistribution}Illustration of how a dataset is distributed over $mn$ cloud owners. The dataset is split into $m$ PPDS (horizontal), with each subset replicated $n$ times (vertical). In practice, the replication can be a function of space, time, required level of consensus,  etc and may vary for each data subset.}
\vspace{-5mm}
\end{figure}


\vspace{2mm}\noindent\emph{Data replication}.  In order to save GPU compute resources, a fraudulent $CO$ can also claim to have trained the model on its subset without actually performing the required training (by supplying a dummy model). To prevent this scenario, our solution also requires that each data subset be replicated (i.e. copied) across multiple $CO$s as illustrated in figure~\ref{figure:datadistribution}. We use the term replicated subsets to refer to data subsets that are replicated. The replicated subsets serve a different function compared to PPDS. The former ensures that model training was indeed performed correctly on each data subset, while the later is used for the purpose of preserving ownership without exposing the full dataset. During federated learning, model training rounds are performed by all the $mn$ COs that hold a subset of a dataset.  Thus, if a fraudulent $CO$ reports an inaccurately trained model, this can be compared against other COs that hold the same replicated subset. As long as the number of fraudulent COs is $< n/2$, any party can verify whether the $CO$ actually trained the model using its data subset or provided a fake model. 



In terms of steps recorded on blockchain, for each dataset, `CreateDSS' is called to have an entry for each of the data chunks in blockchain, i.e. both replicated and PPDS. After each round of training, each $CO$ records a hash of its partially trained model onto the blockchain. However, as the hash values of models trained by $CO$s that hold replicated data subsets would be the same, a fraudulent $CO$ can still read these hash values from the blockchain and report these back again as proof of a trained model. To overcome this problem, each $CO$ appends a random nonce to its partially trained model and then computes the hash. The $CO$ then reports both the hash and the nonce onto the blockchain. Since no two $CO$s can produce the same nonce, the hash reported by each $CO$ is guaranteed to be different. Thus if a fraudulent $CO$ copies any other $CO$'s hash and/or nonce and reports it as its own, then the $CO$ reporting the same hash at a later time on blockchain can be concluded as fraudulent. As long as the number of fraudulent $CO$s is $< n/2$, the correct model corresponding to each data subset will be available via consensus. Additionally, any party that has access to the partially trained model (e.g. $MO$) can append the nonce reported by the $CO$ to the model and compute its hash to verify the hash reported by the $CO$. 

%

\subsection{Training via Federated learning steps on blockchain}

Once a $DO$ shares and verifies the data chunks held by different $CO$s, the corresponding $CO$s are ready to participate in the training process via federated learning. 
The $DO$ publishes details related to its dataset and a contract binding which an $MO$ can lookup these details and use the dataset for training. 
When an $MO$ is convinced about the dataset and the associated contract, it expresses its intent to train an AI model on the dataset. 
The training of models via federated learning proceeds in rounds. Since the dataset is distributed across multiple $CO$s, during each round of training the $MO$ supplies the current running model to all $CO$s. At the end of the training round, the $MO$ receives updated models from all the $CO$s and aggregates these into a single federated model (generally by averaging), which it then supplies to all $CO$s for a new round of training. The $MO$ also decides the termination criteria for training the model, i.e. it evaluates metrics on the federated model to determine if the model requires further training. 


In terms of the sequence of operations recorded on blockchain, models are essentially represented as assets and an $MO$ uses `CreateMod' to create an instance of the Model object on blockchain. The model object holds the following information: Model Owner ID, Model ID, Model Type, Model URL, Training Method and Model Hash. The model type specifies the type of AI model (e.g. DT, Neural Network, etc.,). Model URL points to the current version of the model that is obtained after certain number of rounds of training. 
The training method specifies parameters related to how the training has to be performed by each $CO$
The hash value field holds the hash declared by a $CO$ after a round of training. This is used for verification by the $MO$ when it downloads the trained models for aggregation. 

An $MO$ expresses intent to develop and train a model by creating a $TC$ object in the blockchain using the smart contract `Request TC'.  
It mentions the dataset and the model object while creating the $TC$. 
The `status` field in the $TC$ object tracks the different phases of training. 
A notification is raised by the blockchain when a $TC$ is created, which notifies the $DO$ that a model owner wishes to train on the dataset. The $DO$ then uses the `ApproveTC' transaction to approve the model training by setting the status field to ``APPROVED''. This allows the $DO$ to verify if the same $MO$ has submitted multiple prior requests and approve model training.  An asset called Train Job ($TJ$) is created by the same transaction used by the $DO$ to approve the $TC$ for tracking the progress of individual jobs (training of each of the datasets). Therefore, for each $CI$ there exists a corresponding $TJ$.



\noindent\emph{Randomization of PPDS.~} The training begins with an $MO$ invoking the `StartRound (SR)' smart contract transaction. The $SR$ transaction takes as input the $TC$ and identifies the dataset (and eventually PPDS) that can be used for training the model in that round. While federated learning in general allows training on all the PPDS held by different $CO$s, our solution uses a random set of PPDS during each round of training. For the chosen PPDS all the replicated units are considered. A random selection of PPDS for each round of training ensures that $MO$s cannot deduce any meaningful information about the data held by $CO$s using the partially trained models obtained after each round.  However, training over multiple rounds ensures that all the PPDS are eventually utilized for training. At the end of $SR$, blockchain raises notification for each $CO$ whose PPDS and replicated units has been selected for that round.

The $CO$s listens for events published on the blockchain. If a notification corresponding to the start round transaction arrives with $TC$ ID associated with a $CO$, the corresponding $CO$ participates in a round of training. The $CO$ examines the model object provided in the $TC$ to obtain information about the model and the associated training method. The URL for the training program is encrypted with the keys provided in the training file. The $CO$ downloads the training program and starts training. The trained models are stored in specific object stores. The location of the trained model is signed using the public key provided by MO in training file. The $TJ$ is updated with the hash value of the model and encrypted model location using the smart transaction `UpdateTJ'.  The $rem$ field in the $TC$ is used to track whether all $CO$'s have updated their respective $TJ$s. When the $rem$ matches the total number of PPDS selected, the $MO$ is notified that the training is completed. 


\noindent\emph{Training consensus. } As explained in the previous section, a malicious $CO$ may provide a model that is not accurately trained in order to save GPU resources. The consensus across models trained on replicated data subsets helps avoid this problem. For each PPDS, there exists a set of replication subsets held by other $CO$s. Therefore the $MO$ can check the contents stored in blockchain and verify whether the models trained on replicated subsets yield the same hash values. Incase the hash values are different, then the trained model corresponding to a data subset is obtained via consensus across the respective replicated subsets. This cross validation of models across $CO$s ensures that training on PPDS is valid and that $CO$s do not maliciously declare that they have trained a model without actually training it accurately.





The $MO$ listens for notifications about round completion (i.e. when $rem$ matches the number of PPDS that has updated the $TJ$) and upon receiving downloads the individually trained models uploaded by the $CO$s on cloud object storage. The $MO$ uses its key to decrypt the URL (encrypted by the $CO$ using the key available in the training file that belongs to the $MO$). After downloading the model file, it computes the hash value and compares it with the hash value reported by the $CO$ to ensure that the version the $CO$ claims to have 
uploaded is the same as the one downloaded by $MO$. 

After successfully downloading all the models trained by respective $CO$s, the $MO$ aggregates these into a single federated model and evaluates metrics on the model. Based on the metrics, the $MO$ can invoke the $SR$ again and continue to perform the training process or terminate the training.

\subsection{Trusted Verifiability of Participant actions}
This section shows how the actions of any participant can be verified using transactions stored on the blockchain.


\noindent\textbf {How does MO verify that CO has the dataset that it is claiming to have?}
In our system, a dataset is split across multiple $CO$s and each data subset is also replicated across multiple $CO$s. 
Thus multiple $CI$s exist for the same data subset. When a data subset is used for training, all copies of the same data subset is used 
for training (albeit in different $CO$). Each $CI$ will have the hash value of its data subset. 
A consensus is used to check if the result reported by the $CO$ owning the same subset are same. 
 %
For example, consider a data subset replicated across $5$ $CO$s. Thus all $CO$s will train on the same data subset. 
A consensus on correct training can be set to verify the output from at least two $CO$s.  The $MO$ can wait 
for the response from $2$ $CO$s and verify the model. If the models match, then the data subset that $CO$ claims to have is correct.

\noindent\textbf{How can we ensure that 2 COs having same data subset do not copy each others hash?}
In the setting where a data subset is replicated among multiple $CO$s, one $CO$ can copy the hash of a data subset to its $CI$ and falsely claim to have a copy of the data subset. 
To avoid this, each $CO$ computes the hash on the data subset appended by a nonce, before publishing its hash. 
It also publishes the nonce used in the $CI$. 
Therefore a $DO$ can verify if any two $CO$s reported the same hash or not. 

\noindent\textbf{How does CO ensure that DO has given it the correct dataset?}

In order to establish that it has not received a bad data subset or an in-correct one (in case a $DO$ colludes
 with another $CO$), a $CO$ can verify the hashes uploaded by other COs that have the same subset.
 To identify same subsets, the $CO$ computes the hash of the data it has and compare it with others. 
  It adds the nonce declared by another $CO$ to its data subset and recalculates the hash on its subset. 
  If the hash matches with the one uploaded by the $CO$ whose nonce was used,
  then it is likely that both COs were given the same subset of the dataset.
  Thus, a $CO$ can cross verify whether it has received the correct data subset from a $DO$.

\noindent\textbf{How does DO verify that CO has got the correct dataset?}
The $CO$ upon receiving a data subset updates the $CI$ with the hash of its subset along with the nonce information.
 $DO$ recomputes the hash along with the nonce of the $CO$ and verifies if both match and sets the status to `VERIFY'.
 Only the $DO$ is provided permission to update the $CI$.
 
 \noindent\textbf{ How does an MO ensure that the partially trained model uploaded by the CO is the same that it has downloaded?}
 A $CO$ trains the model received from a $MO$ on its data subset. Once the training is over, the $CO$ updates the $TJ$  with the hash of the model.  
Since several $CO$s posses the same data subset,  a malicious $CO$ can copy the hash value of the model provided by a $CO$ that actually spends resources to train a model. 
 In order to avoid this, a nonce is added by a $CO$ to the model before computing its hash. 
When the model url becomes available, the $MO$ downloads the model and computes the hash by appending the nonce provided in the $TJ$.
It the hashes match, then the model uploaded by the $CO$ is the same as the one it has downloaded. 

\noindent\textbf {How does MO ensure that CO simply doesn't copy the hash from another CO having same data subset?}
A $CO$ adds a nonce to model to compute the hash and no two $CO$s can declare the same nonce.
Therefore if a $CO$ copies the hash computed by another $CO$, then $MO$ can recompute the hash by using the model uploaded by $CO$ and the nonce declared by it. 
If the hashes match then the model uploaded by the $CO$ is same as the one it has declared.

\noindent\textbf{How does DO ensure that the MO doesn't control which data subsets will be used in a training round?}
The data distributed to the $CO$s is recorded in the blockchain through the $CI$. 
In this case, an $MO$ can easily query the blockchain and get the list of $CI$s and attempt to identify the distribution of the data subsets.  
In our system, the selection of $CO$s to execute model training in each round is performed by the blockchain through a chaincode function `start Round'.
 Although invoked by the $MO$, this chaincode function which holds the logic of picking up a random subset of $CO$s to conduct a round of training, is carried out by the underlying blockchain which is completely agnostic of the $MO$. This is also one of the advantages of using a blockchain.

\noindent\textbf{How does DO ensure that MO cannot access the data shared with CO?}
The $CO$s hold the data subsets and can therefore collude with a $MO$ to share the data. 
If a subset of $CO$s collude with the $MO$, the $MO$ can gain access to a part of the dataset, in which case the data may be compromised and the ownership is lost. 
However this is not possible since the $MO$ only has access to the blockchain and does not know the identity of any $CO$s. Each $CO$ is effectively represented as a random hash value in the blockchain. Therefore, an $MO$ does not obtain access to the actual identify or location of the $CO$. 

\section{Experiments, Observation and Results}
\label{sec:experiments}

We have implemented our system - 15 transactions in all, using chain code functions. Using these transactions
one can perform all operations of the protocol and complete model training activities. We have evaluated the protocol
for two important metrics: latency and throughput, typical of distributed systems that require scalability.
Our experiments were performed with permissioned blockchain network 
(Hypledger fabric version 1.2.0-rc1) components deployed
as Docker containers running atop Soft-Layer \cite{softlayer} servers. Each component
was provisioned a separate server with 32 cores, 64GB RAM
and ran Ubuntu16.04. We use Hyperledger Caliper (or Caliper) \cite{caliper} as
the benchmarking tool. Caliper allows users to measure the performance
of a blockchain implementation with a set of predefined use
cases and produces reports containing a number of performance indicators,
such as tps (Transactions Per Second), transaction latency,
etc.

All our experiments use standard default settings of configuration parameters that comes
with the Fabric. This is to emulate the general behavior of the blockchain and one can expect better responses 
in optimized settings. The block size for all our experiments was 500
and the block timeout was 1s. The default block formation policy
was considered as 2:3:1. The transaction submission rate
were 500tps,1000tps and 1500tps. We have tried with other higher transaction
rates, but the metrics felt sharply down. We ran each experiment 30 times and in each run a
total of 100,000 transactions were submitted. We report the average
across all the runs.

\begin{figure}
  \centering
  \includegraphics[scale=0.3] {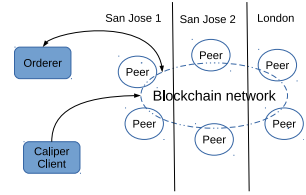}
  \caption{Ilustration of a blockchain network used to study performance of our protocol. The number of peers shown in the picture is only for illustration and in reality 
  is scaled upto 24 in each location.\label{fig:network-setup}}
\end{figure}

Peers in the blockchain were setup in different locations to faithfully reproduce experiments
closer to real world scenarios.  We had considered upto 24 peers in each location and 2 data 
center (DC) locations viz.,
San Jose and London. Experiments were conducted with nodes located within
  a Data center (Single DC)
or across data center locations (2 DC setup). In the 2 DC setup we have two configurations.
One where the geo locations were San Jose 1 and San Jose 2 and another where peers are distributed between
between San Jose 1 and London geographies.
The number of peers were equally distributed
between two locations. For example, in a 4 peer network 2 DC setup- 2 peers are located 
in San Jose location and 2 peers are located in London.  Figure \ref{fig:network-setup} illustrates the network setup used in our experiments in different locations. We have one orderer 
and one client both in the San Jose location.

\subsection{Effect of Increasing number of Peers}

\begin{figure}[h!]
  \includegraphics[scale=0.25]{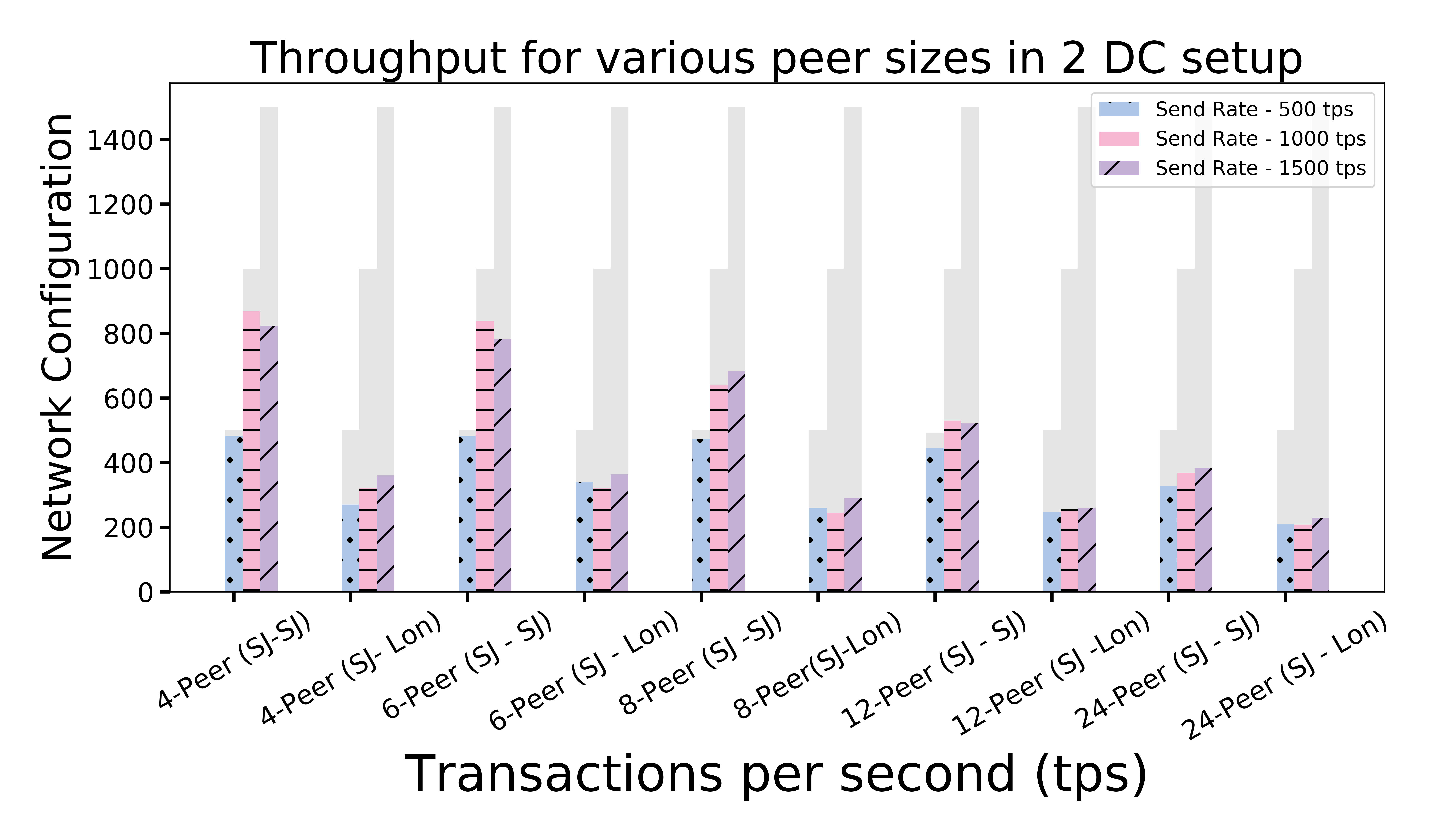}
  \caption{Effect on Throughput with increasing peer sizes distributed
  across 2 geographically distributed data centers.\label{figure:tx_send}}
  \vspace{-0.2in}
\end{figure}

\begin{figure}[h!]
  \includegraphics[scale=0.2]{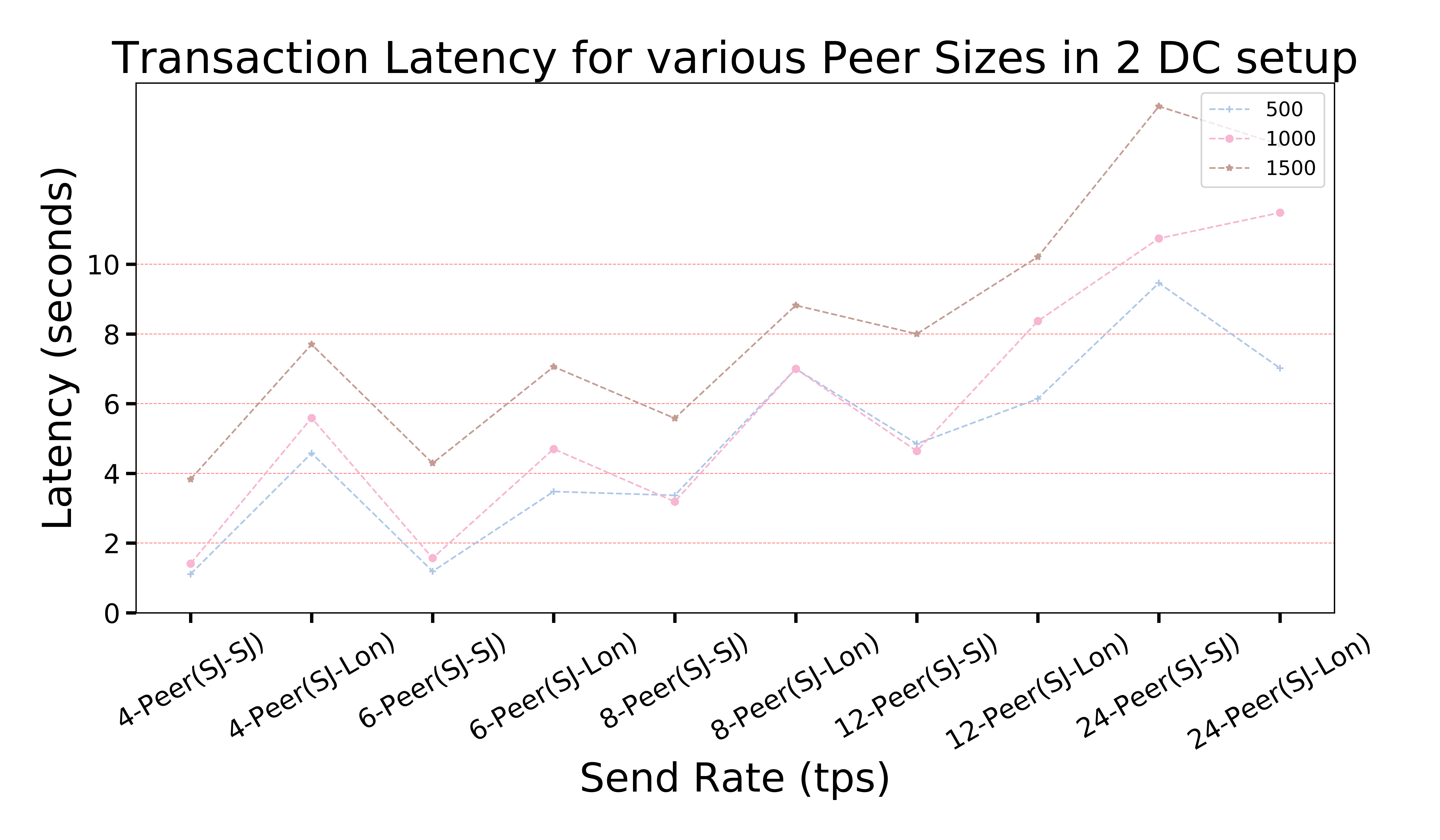}
  \caption{Effect on Latency with increasing number of peers distributed across 2 geographically
  distributed data centers \label{figure:lt_peer}}
  \vspace{-0.2in}
\end{figure}

We study the effect of increasing peers in a single DC and 2 DC setup. 
Figure \ref{figure:tx_send} and Figure \ref{figure:lt_peer} shows the different 
transactions per second and latencies 
observed with increasing number 
of peers in a 2 DC setup.   

Our experimental results indicates that the throughput falls by about $1.5$ times when the number of peers
are doubled. For instance, increasing the peers from $4$ to $6$ reduces
 the average throughput by about $30$\% compared to a $4$-peer configuration. While this is expected due to the 
 endorsement times and varying queue sizes, 
 the result
 is also due to the large ping latencies observed 
 acorss the machines in these different locations.

Latency is measured end-to-end, from the time of  submitting a call to the blockchain 
to the time it was committed to the ledger by the peers. 
Figure \ref{figure:lt_peer} shows the average latencies 
computed for different peer configurations.
With increase in peer size spread across even 2 different locations 
the latency increases by about $4$-$6$ times.
Our measurements of the ping times between the client and different peers 
 in different locations show that the ping latencies
vary in the range of $300$ms to $3$s.  As blockchain peers 
have to communicate with the client during the endorsement
 and  commit phases,  communication costs play a significant role in latencies on blockchain.
 With high network speeds one can expect the latencies to reduce, however latency has  a significant 
 impact on the throughput.

\subsection{Effect of Send Rates}
We measure the effects of varying Send Rates. Send Rate emulates 
different loads that each peer gets from their clients.
We experiment with three different send rate $500$tps, $1000$tps, $1500$tps and observe the
throughput and latencies of the blockchain system. We have tried experiments with higher 
send rates but the throughput falls sharply and therefore pivot our observations
around these numbers.

\begin{figure}[h!]
  \includegraphics[scale=0.25]{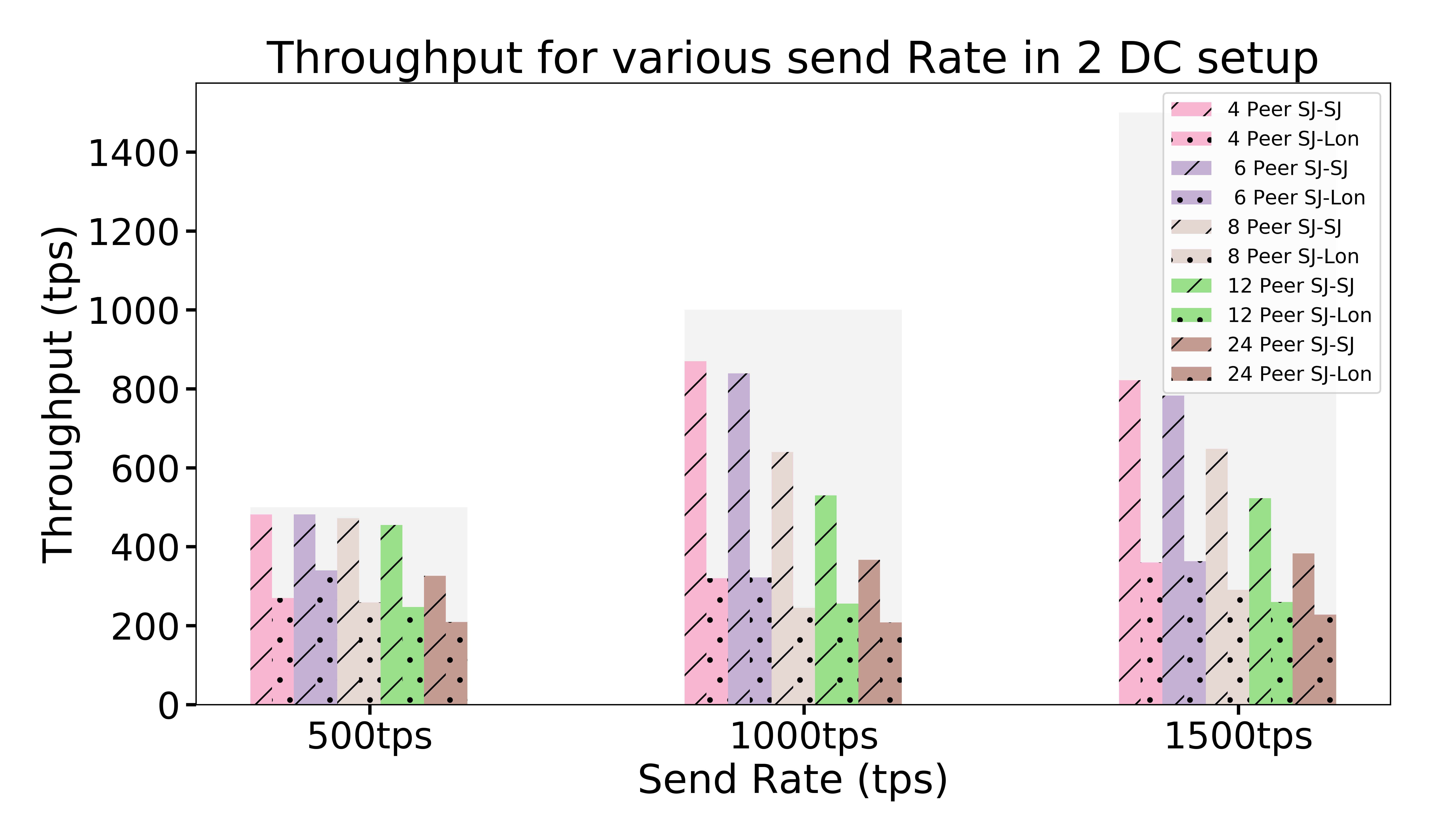}
  \caption{Effect on Throughput with increasing  send rates of transactions\label{figure:send_rate_throughput}}
  \vspace{-0.2in}
  \end{figure}

  \begin{figure}[h!]
    \includegraphics[scale=0.25]{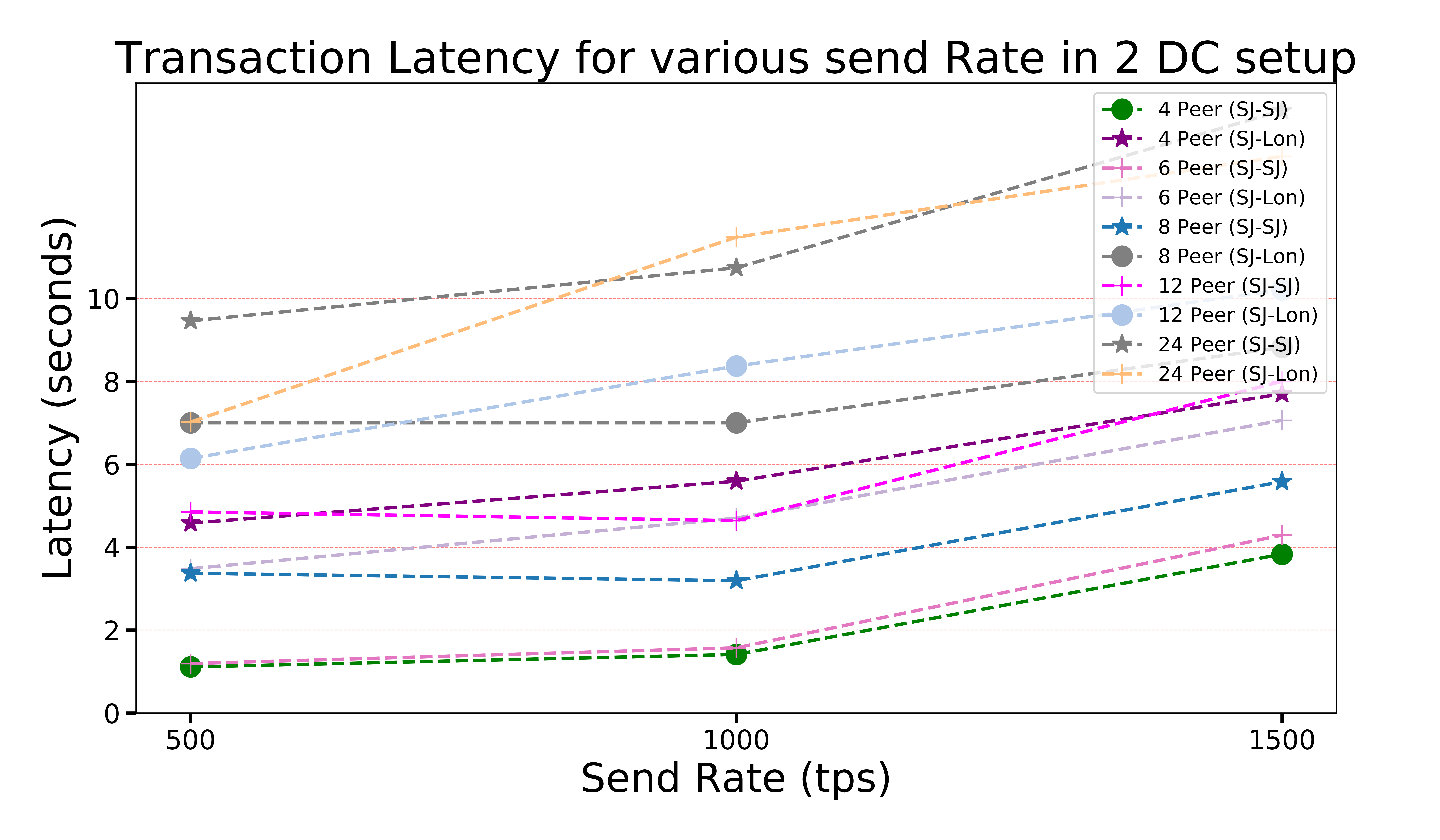}
    \caption{Effect on Latency with increasing send rates of transactions  \label{figure:send_rate_latency}}
    \vspace{-0.2in}
    \end{figure}
 
Figure \ref{figure:send_rate_throughput} shows the observed throughput averaged over all transactions
for varying peer sizes.  As our objective was 
to measure the `limit' of the blockchain to handle incoming rates (i.e. incoming rate handled by blockchain without drop in throughput),
we average the rate over $30$ different rounds for all transactions.  In the Figure `SR-TP' shows the 
gap between the throughput and the `send rate'. 
It can be seen that the blockchain can scale effortlessly
for send rate of $1000$tps and falls sharply for $1500$tps. 
The sharp fall is due to the fact that as peers
increase, time to endorse and commit also increases significantly reducing the throughput rates.

Given these throughput rates we try to
answer the question about the number of model trainings that our system  can support.  
Considering a single $DO$, $CO$ and $MO$ training a data set it takes about $15$ transactions to complete the 
entire model training. It does not include the model training time which is done by
the cloud owners and not by the blockchain.  
On average about $950$ transactions are supported ($4$ DC being the best) for `send rate' of $1000$.      
This implies that about $65$ model trainings per second could be supported easily without additional optimization. Compared to both 
 existing networks like Bitcoin \cite{bitcoin} that supports roughly $7$tps and number of models 
reported  by current online systems \cite {kaggle} (about $200$ models a month) our system
 is well equipped to scale for large scale model training.


  Figure \ref{figure:send_rate_latency} shows the average latency (for all transactions)
for different send rates. Latency increases sharply when the transaction `send rate'
 is more than
$1000$ tps. This is due to increase in number of transactions within the blockchain.
The blockchain components (such as orderer and endorsers) slow down,  spending their time
in book keeping of records. Further, the queue sizes for the orderer increase 
as the transaction input rate increases. Transactions 
spend time in the kafka queue (queue within the orderer
of blockchain that stores the transactions for making blocks)
 before getting committed. This blockchain setup can support approximately $1000$ tps after which the latency increases sharply.

\subsection{Transactionwise comparisons}
\begin{figure*}[h!]
  \begin{subfigure}{\textwidth}
    \centering
    \includegraphics[scale=0.25]{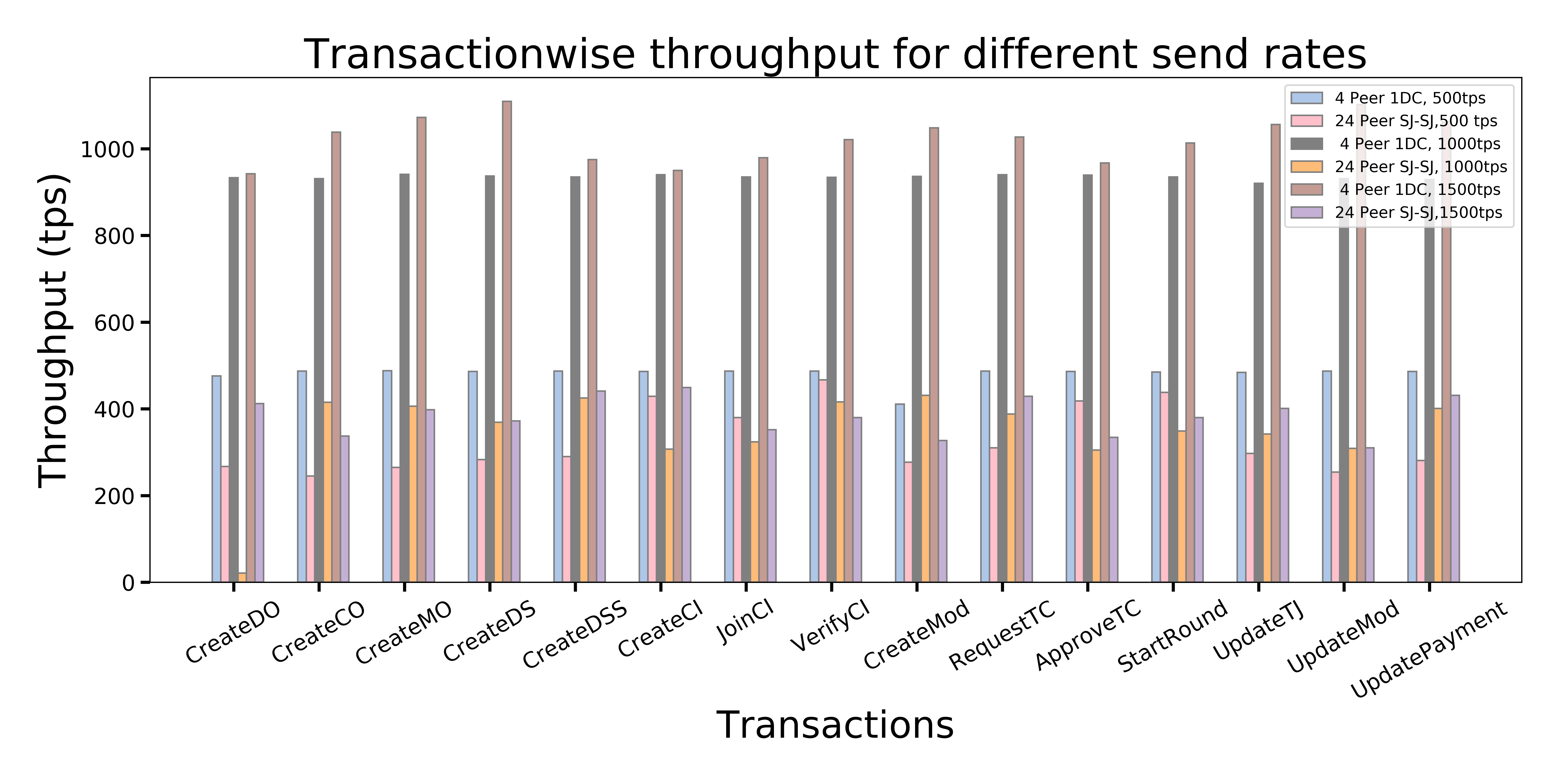}
    \caption{Transactionwise Throughput scaling for select peer sizes and send rates}
    \label{fig:sfig1}
  \end{subfigure}
  \\
  \begin{subfigure}{\textwidth}
    \centering
    \includegraphics[scale=0.25]{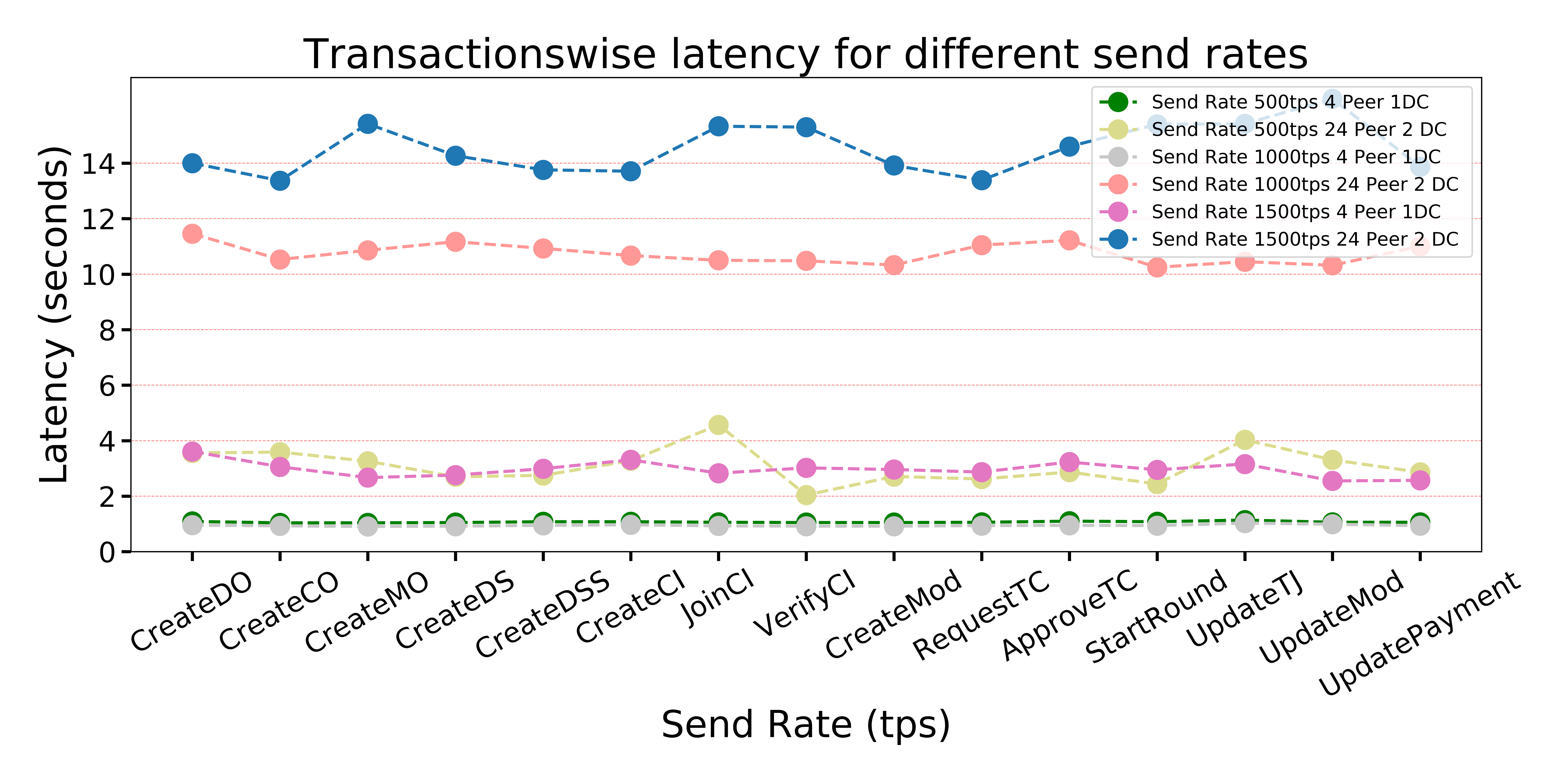}
    \caption{Transactionwise Latency for select peer sizes and send rates}
    \label{fig:sfig2}
  \end{subfigure}
  \caption{Throughput and Latency impact for different transactions\label{fig:transactionwise}}
  \vspace{-0.2in}
\end{figure*}

We study transaction wise impact due to increasing peers and different send rates. 
Figure \ref{fig:transactionwise} (a)  shows the different throughput for each smart
 contract in the solution. Similarly, latency for all transactions across two different network settings ($1$-DC and $2$-DC) is presented in
 Figure \ref{fig:transactionwise} (b).

 All smart contracts follow the similar trend
  in the number of transactions observed for a particular peer distribution. 
  The best observed case is one of $4$ peers in a single DC which
 can support up to $1000$(tps).
The throughput reduces significantly for the $24$ peer network spread across $2$ DCs.
As mentioned earlier
this reduction is due to the ping times observed between the peers and the 
client. In short, 
individual
transactions have similar throughput or latencies. This is key to the design as no single API
 would block or create undue delays in the transaction execution leading to a bottleneck. 

The average time taken to 
finish all the $15$ transactions is  about $15$s in a 4 peer $1$-DC setup. Thus, for a model trained the total time
spent on the blockchain is roughly $15$ seconds. This time, compared to the actual model training
time which can span hundreds of minutes is insignificant compared to the benefits obtained. 

\subsection{Comparison to centralized transaction system}

We have also performed experiments considering an alternative centralized 
system that can provide similar functionality using a traditional database.
Operations in the blockchain are mapped to either insert operations (where 
information is recorded) and procedural sql (where logic is available) in database.
For the experiments, we use single node open source database Postgres \cite{postgres} setup and
 use Pgbench \cite{pgbench} to perform the experiments. Pgbench reports about $2000$-$3600$ tps 
 for a send rate of $100$K transactions using $8$ clients and each client generating workload using 8 threads.
 The latency as observed using 
 Pgbench ranges from 
 $700$-$3000$ ms (including the connection time).
 These observations are on a highly
optimized single node transaction systems and accommodates
roughly twice the number of transactions than the proposed blockchain system. However,
trust is left to the external intermediaries.
Our system at a reasonable drop in transactions compared to 
well optimized Postgres provides
inherent trust and outweighs the costs involved.

In summary, we see that the proposed protocol can scale efficiently for training large number of 
AI applications (supporting $1000$tps in a $4$-DC setup, equivalent to $65$-$70$ models trainings per second).  A model training using our system
introduces approximately $15$s of delay which remains insignificant compared to the 
large time scales involved in training AI applications. Even when compared to traditional 
database systems that provide similar functionality, the performance levels are comparable, thus enabling an affordable system that ensures privacy and ownership of AI assets in an otherwise trustless environment.
   

\section{Related work}
\label{sec:relatedwork}

Our system  can be contrasted against prior frameworks 
for AI marketplaces along the dimensions of design, privacy and data leaks.

Kaggle \cite{kaggle} is one of the earliest centralized ventures, which provides a marketplace platform for data owners and model developers to collaborate. However both data and models are publicly available to participants.
In our distributed design, data and models are maintained by individual stakeholders while blockchain ensures transparent execution of training of models on data. Moreover the design relieves blockchain peers from performing AI training related operations, which enables plug-and-play of our system over any blockchain network.

SkyChain~\cite{SkyChain} solution is specific to medical AI services. 
Both model and datasets are uploaded to the SkyChain database and SkyChain provides the infrastructure to train the model. 
Data and models being core AI assets in the marketplace, one would expect their value and monetary benefits to grow with every usage. 
\cite{Datum}  is a blockchain based decentralized database for storing personal data in encrypted fashion. Buyers can pay DAT 
tokens to buy the "Keys" to the datasets.  However, these marketplaces expose either the data or model or both to the participants, resulting in loss of value over time.  
Ocean protocol \cite{ocean}  is another blockchain based AI marketplace and uses a reputation based system to remove fake Data and dishonest participants for the systems. 
Our system can operate across multiple administrative domains providing complete privacy and ownership of AI assets. With every application using the data or model,
our platform has the potential to provide monetary benefits to all stakeholders. 
Droplet \cite{droplet} operates using token transfer. Data is encrypted using symmetric encryption. However the data is exposed to services that perform training and is therefore prone to leakages.

Closest to our work is the Danku protocol \cite{algorithmia} proposed by Algorithmia, 
which enables the operation of an AI marketplace wherein blockchain peers are utilized for both training and data storage. 
However, multiple peers execute the training of the same model by downloading data, leading to redundant computation and storage utilization. 
OpenMined \cite{openmined}
focusses on exposing data from end users who own very small amount of data. Our solution allows
large data sets to be trained without relying on the data owners to provide 
computational resources. Computable labs \cite{computable-labs} is yet another AI marketplace
that trains models, however the data is completely exposed to the buyers who can potentially
leak the data to other buyers. Our approach 
 employs lightweight solution where the assets are stored off-chain.
 The blockchain peers are not overloaded with the training or storage tasks. 
 In \cite{algorithmia}, with the increase in the number of training jobs, peers would spend significant amount of time in training, which renders it inefficient for large data sets or training that requires longer durations.  Our unique proposition is to guarantee complete ownership of data and models without exposing them to any participants in the system. 

\section{Conclusions}
\label{sec:conclusions}

Lack of data and model privacy leading to the subsequent loss of value and ownership have impeded the growth of both centralized and decentralized AI marketplaces. 
 We present a novel mechanism for protecting
the privacy and ownership of assets in a decentralized and trustless AI marketplace using blockchain.
Our system chaincode functions are
 set up to incentivize all participants to truthfully 
 record their actions on the distributed ledger, so 
 that the underlying blockchain system holds verifiable evidence 
 of expected behavior, wrongdoing and dispute resolution. Our implementation 
  using the Hyperledger Fabric shows that our system
   can support large scale model training and provides a viable
    alternative to centralized AI systems that
     do not guarantee data or model privacy. 

In future, we intend to include more comprehensive algorithms for homomorphic encryptions of models and utilizing encrypted models in federated learning. We are also
working towards an evaluation mechanism where the stakeholders can transparently evaluate
the models and data and get paid for their contributions. 
\bibliographystyle{IEEEtran}
\bibliography{ref}

\end{document}